\documentclass[twocolumn,secnumarabic,amssymb, nobibnotes, aps, prd,superscriptaddress]{revtex4-1}

\usepackage{graphicx}% Include figure files
\usepackage{dcolumn}% Align table columns on decimal point
\usepackage{bm}% bold math
\begin{document}

\preprint{CoCu XMCD and XAS manuscript}

\title{Emerging Magnetic Order In Copper Induced By Proximity To Cobalt: A Detailed Soft X-Ray Spectroscopy Study}

\author{Zhao Chen}%
 \email{Email: zhaochen@slac.stanford.edu}
\affiliation{Department of Physics, Stanford University, Stanford, California, 94305 USA}
\author{Hendrik Ohldag}
\affiliation{SLAC National Accelerator Laboratory, 2575 Sand Hill Road, Menlo Park, California, 94025 USA}
\author{Tyler Chase}
\affiliation{Department of Applied Physics, Stanford University, Stanford, California, 94305 USA}
\author{Sohrab Sani}
\affiliation{Department of Physics, New York University, New York City, New York, 10003 USA}
\author{Roopali Kukreja}
\affiliation{Department of Materials Science and Engineering, University of California Davis, Davis, California, 95616 USA}
\author{Stefano Bonetti}
\affiliation{Department of Physics, Stockholm University, Stockholm, 10691 Sweden}
\author{Andrew D. Kent}
\affiliation{Department of Physics, New York University, New York City, New York, 10003 USA}
\author{Eric E. Fullerton}
\affiliation{Center for Memory and Recording Research, University of California San Diego, La Jolla, California 92093 USA}
\author{Hermann A. D\"{u}rr}
\affiliation{SLAC National Accelerator Laboratory, 2575 Sand Hill Road, Menlo Park, California, 94025 USA}
\author{Joachim St\"{o}hr}
\affiliation{SLAC National Accelerator Laboratory, 2575 Sand Hill Road, Menlo Park, California, 94025 USA}

\date{\today}% It is always \today, today,

\begin{abstract}
We present an x-ray magnetic dichroism (XMCD) and soft x-ray absorption spectroscopy (XAS) study to address the nature of emerging magnetic order in metallic Copper as Cobalt is added to the matrix. For this purpose line shape and energy position of XAS and XMCD spectra will be analyzed for a series of Co/Cu alloys as well as a multilayer reference. We observe an increased hybridization between Cu and Co sites as well as increased localization of the Cu d-electrons and an induced magnetic moment in Cu. The emergence of long range magnetic order in non-magnetic materials that are in proximity to a ferromagnet is significant for a comprehensive interpretation of transport phenomena at ferromagnetic/non-magnetic interfaces, like e.g. the giant magnetoresistance effect. The presented results will further enable us to interpret Cu XMCD and XAS spectra acquired from unknown Co/Cu samples to identify the environment of Cu atoms exhibiting proximity induced magnetism.

%We observe the emergence of localized, magnetic 3\textit{d} Cu states in Co/Cu alloys using high-resolution X-Ray Absorption Spectroscopy (XAS) and X-Ray Magnetic Circular Dichorism (XMCD) spectroscopy at the Copper L$_3$ resonance. Increased Co concentration in the alloy leads to two prominent spectroscopic effects consistent with formation of these localized Cu states: (1) energy shifts of the Cu L$_3$ XAS peak by 0.5eV relative to the Cu L$_3$ XMCD peak, resulting in a shared maxima with respect to energy in the limit of pure Co, and (2) narrowing of the XAS peak proportional to Co concentration. We compare our results to XAS spectra taken of multilayered samples, and show that our alloyed Cu atoms behave analogously to Cu atoms near a Cu/Co interface. Our results thus provide novel, key insight into the behavior of Cu when placed near a ferromagnetic interface, which is crucial for modern spintronics research where Cu is not only often used as a spacer in F/N/F multilayer devices, but also is often fabricated to such low thicknesses that interfacial effects begin to dominate the physics.

\begin{description}
\item[PACS numbers]
\verb+75.70.-i+, \verb+75.50-Cc+, \verb+78.70.-Dm+
\end{description}
\end{abstract}

\pacs{Valid PACS appear here}% PACS, the Physics and Astronomy
                             % Classification Scheme.
%\keywords{Suggested keywords}%Use showkeys class option if keyword
                              %display desired
\maketitle

%\tableofcontents
\section{\label{sec:level1}Introduction}

Transport of spins across interfaces is becoming increasingly pertinent in understanding important magnetic phenomena such as the spin hall effect \cite{sh1,sh2,sh3} and ultrafast demagnetization \cite{superdiffusive1,superdiffusive2,spinflip1,spinflip2}. From an engineering perspective, such effects will also begin to dominate the physics as state-of-the-art magnetic devices become smaller. For instance, interfaces have long been known to be strong scattering centers in tunable systems such as magnetic tunnel junctions and GMR devices~\cite{parkinPRL,zhangPRL,zhangPRB,yuasaSCIENCE}. Surprisingly strong coupling/hybridization between Cu and nearby ferromagnetic atoms has also already been observed~\cite{samantPRL,tersoffPRB,karisPRB,pizziniPRL}. Furthering this narrative, it was recently demonstrated in~\cite{roopaliPRL} that spin scattering of a spin-polarized current carried into Cu from an adjacent Co layer was surprisingly two orders of magnitude stronger at the interfacial Cu atoms. Interfacial physics will thus play an increasingly crucial role in our understanding of magnetic dynamics at the nanoscale, which motivates our current study: to spectroscopically study copper (a non-magnet) as it is put increasingly close to magnetized cobalt atoms. Such a situation is commonly found in practice; copper is often used as a non-magnetic spacer layer material or electrical conductor in magnetic devices, and as a consequence is often in close proximity to ferromagnetic atoms at interfaces or in an alloy. Here we will present a comprehensive soft x-ray absorption spectroscopy (XAS) and magnetic dichroism (XMCD) study of Co/Cu alloys. We choose to focus our efforts on alloyed samples as their atoms exist in more uniform chemical environments and they thus produce more interpretable data. They are also more straightforward to fabricate and tune to the desired concentrations. We use soft x-ray dichroism absorption spectroscopy at a synchrotron source as such a measurement provides an element specific magnetic probe \cite{stohrMagnetism}.\\
\indent In this article, we will show that the proximity of non-magnetic Cu atoms next to magnetized Co leads to two main effects in the Cu XAS/XMCD spectra. First, we see a core-level red shift in the Cu XAS threshold, analogous to those observed by Grioni et al. and Nilsson et al. in~\cite{grioniPRB,nilssonPRB}. But we also observe an opposite shift in the visible XMCD Cu feature, resulting in a blue-shift of the magnetic Cu peak with increasing Co concentration. In the limit of high Co concentration (i.e. dilute Cu), the Cu magnetic signal is energetically degenerate with respect to the Cu absorption edge maximum. This is the result of Co imprinting its own spectroscopic features onto Cu atoms. Second, we will show that the FWHM of the Cu XAS resonance decreases considerably with increasing Co concentration. Peak narrowing has been observed in one Co/Cu system before~\cite{nilssonPRB}, but we will show a steady linear trend within our range of concentrations of the FWHM of the Cu XAS peak, and show a very dramatic narrowing in the limit of high Co. These observations are consistent with the interpretation that the onset of magnetization and localized d-character in the Cu DOS is a direct consequence of proximity and hybridization with nearby Co atoms. %Last, we discuss the satellite peaks on the blue side of the Cu L$_3$ XAS peak, which are primarily reflective of 2p$_{3/2}\rightarrow$ 4s transitions, and demonstrate that these satellite peaks coalesce in both the XAS and XMCD channels with more Co concentration. These satellite peak shifts also result from Co imprinting itself onto nearby Cu atoms and the latter taking on increasing Co character. All trends in our data thus indicate that with increasing presence of Co, there is a smooth increase in Cu d-d hybridization and a smooth emergence of localized magnetic states in Cu with high amounts of $d$ character.

\section{\label{sec:level1}Experiment}

\indent Our experimental results were obtained using soft x-Ray absorption spectroscopy (XAS). XAS entails the measurement of the ratio of x-ray photons transmitted through matter, which allows us to directly calculate the absorption cross section. By tuning the photon energy such that core-level electrons of a particular elemental species are excited into available empty valence states above the Fermi level one can probe the unoccupied density of states with elemental specificity. For our samples, absorption at the L$_3$ resonance corresponds to transitions from 2p$_{3/2}$ states to the valence 3$d$ or 4$s$ states of either Cu or Co. By using circularly polarized x-rays, core-level electrons with a particular spin are excited preferentially, and since optical transitions conserve angular momentum one obtains a element-specific and spin-resolved probe, i.e X-ray magnetic circular dichroism (XMCD). XMCD probes an absorption difference between magnetizations parallel and antiparallel to the beam propagation direction when incident photons are circularly polarized. The intensity difference can be shown to be proportional to the component of magnetization in the beam propagation direction~\cite{tholeXMCD, stohrMagnetism}. In this paper we report most data as $\mu^{(XMCD)} = \mu^{(+)}-\mu^{(-)}$, the absorption coefficient difference for the two different polarities of incident circular polarization. \\
To ensure the robustness of our conclusions, Co/Cu alloy samples were fabricated using two separate magnetron sputtering deposition systems at New York University (NYU) and the University of California, San Diego (UCSD). A Co/Cu multilayer sample with structure Pt(50)[Co(10)Cu(4)]x55Ta(50) layers was also fabricated at UCSD for comparison to alloy data. All samples were grown on 200nm thick SiN membranes to allow transmission of X-Rays, and all alloyed samples grown were both seeded and capped with Pt (3nm at NYU, 5nm at UCSD). Sputtering rates were less than 1\AA/s. Nominal Cu concentrations for alloyed samples were 10\%, 50\%, 90\%, and 100\% (reference sample), but accurate concentrations were determined by cross-referencing the XAS spectrum signal amplitude with standard tabulated L-edge absorption cross-sections for Cu metal (for example, see \cite{stohrMagnetism}). \\
\indent Measurements were performed at the Stanford Synchrotron Radiation Lightsource (SSRL), using the XMCD setup at Beamline 13-1. The beamline provides x-rays via an elliptically polarized undulator, emitting circularly polarized photons with energies between 250eV to 1250eV. Most measured X-ray absorption spectra focus on the L$_3$ and L$_2$ edge of Cu around 930eV to 950eV. An upstream monochromator provides energy resolution of $\approx 0.1$eV at this photon energy. To account for photon energy drifts due to varying thermal loads on the x-ray optics, XAS of reference samples were taken in regular intervals. This allowed us to achieve an accuracy of the relative energy position of $\pm 0.05$eV. Transmitted x-ray photons are detected in a large-area photodiode downstream of the sample, and transmission measurements are normalized to a photon intensity signal extracted via electron yield on an upstream gold grid. XMCD spectra were taken at an angle of incidence of 45 degree with a magnetic field of $\pm 0.25$~T applied along the direction of the x-rays. The sign of the external magnetic field was reversed at each energy point in the spectrum before moving to the next photon energy value. This allows us to measure both $(+)$ and $(-)$ spectra simultaneously, which minimizes potential issues with long-term beam drift. Reference spectra on a pure Cu sample were recorded at the beginning of each day of measurement to recalibrate any energy shifts that may have accumulated overnight.

\section{\label{sec:level1}Results and Discussion}

\begin{figure}
\centering
\includegraphics[width=0.53\textwidth]{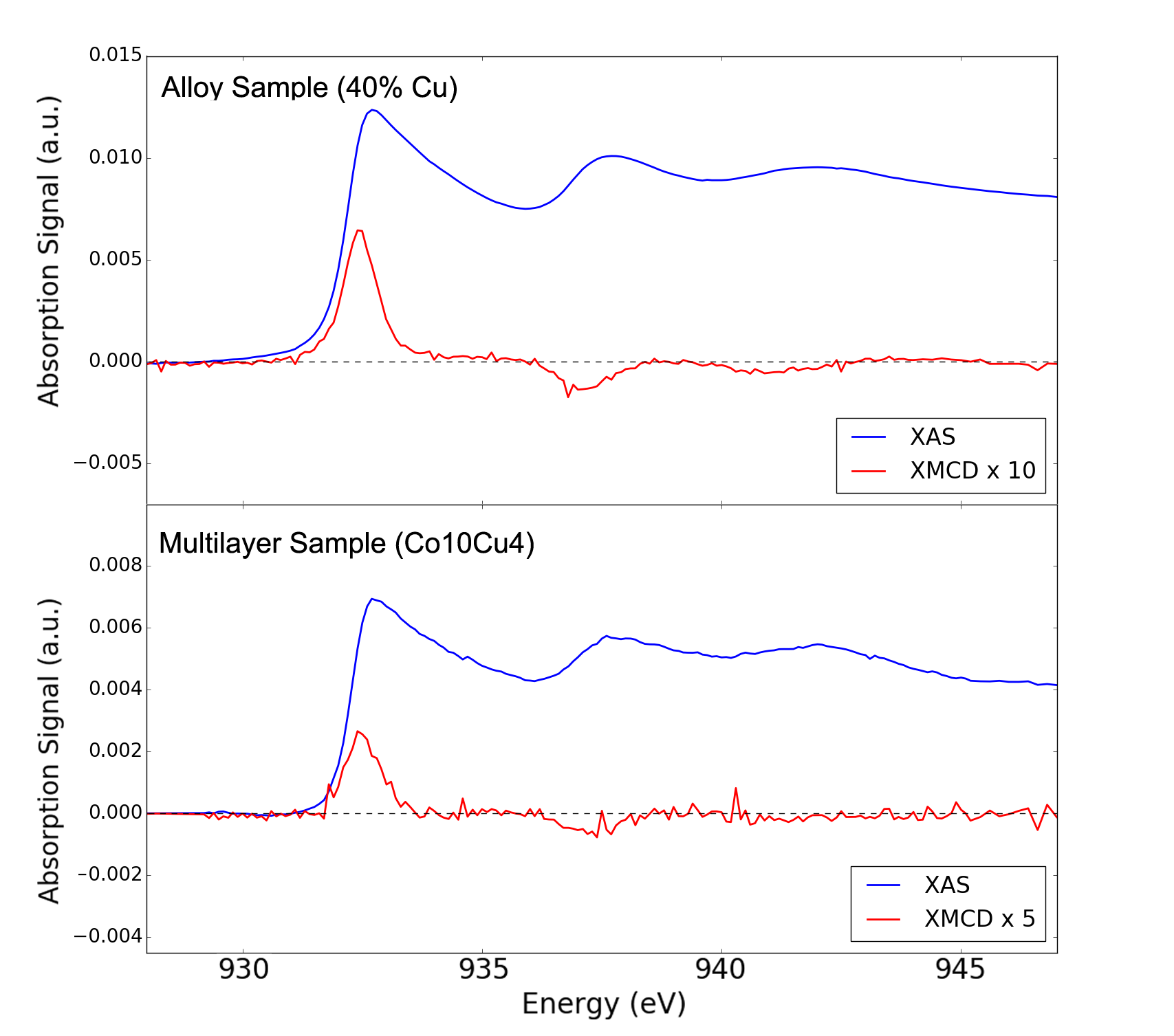}
\caption{\label{fig:1} (color online) High-resolution XAS and XMCD (x5) spectra across the L$_3$ edges of Cu for a alloyed samples containing 40\% Cu (top), and comparison with a [Co(10)Cu(4)]x55 multilayer sample (bottom). All spectra are normalized to a gold grid $I_0$ signal, and the XMCD spectra were inverted for easier comparison of peak positions. A dotted line at zero is provided for reference. XMCD spectra are provided as scatter plots for clarity, due to the intrinsically larger noise in Cu XMCD spectroscopy.}
\end{figure}

\subsection{XMCD Spectra}
Figure~\ref{fig:1} shows the full Cu spectrum across the $L_3$ edge, for both a 40\% Cu alloy sample and a [Co(10)Cu(4)]x55 multilayer. These samples were chosen because their nominal relative Cu concentrations are similar, and hence should serve as a meaningful comparison between multilayer and alloyed spectra. The XMCD spectra presented are inverted relative to the usual convention for easier comparison of the main XMCD/XAS L$_3$ peaks. Values of the absorption coefficient are presented as an average $\mu^{(XAS)} = \frac{1}{2}(\mu^{(+)}+\mu^{(-)})$ of the measurements at opposite magnet polarities, while the XMCD coefficients are presented as a difference $\Delta \mu^{(XMCD)} = \mu^{(+)}-\mu^{(-)}$ of these two measurements. Coefficient values were calculated from measured data assuming 30nm total thickness and with Cu percentage in the alloy calculated by comparing the XAS amplitude with known Cu L$_3$ cross-sections \footnote{One potential issue with using the XAS peak as a way to measure Cu percentage in our alloy samples is that the XAS amplitude will increase not only with more Cu in the sample, but also with the increased 3$d$ empty states that result from increased Co percentage~\cite{nilssonPRB}. However, we note that the absorption effect is much more significant than the effect from increased 3$d$ states, which does not exceed a $25\%$ absorption difference even for very high $Co$ concentration. Hence, using absorption as a thickness calibration is still robust enough to be valid.}.\\
\indent The multilayer spectra are presented to show that alloyed samples exhibit similar physics to those of multilayer samples, which allows us to generalize our conclusions from alloy measurements to multilayers as well. As shown in Figure~\ref{fig:1}, there are three peak features per spectra. We observe a very close correspondence between the peak features in both the alloyed and multilayer spectra; all peak position differences lie within our machine's resolution of 0.1eV \footnote{We note that the XMCD signal in the multilayer sample appears to be stronger relative to the XAS signal; this is reflective of the slightly lower concentration of Cu in the multilayer (29\% versus 40\%), which increases the relative amount of induced magnetism in the multilayer. The increased relative XMCD amplitude is quantitatively consistent with the trend presented in the inset of Figure~\ref{fig:4}.}.\\
\indent We also observe auxiliary peaks at 937.0eV and 941.0eV in the spectra, which show an inverted magnetic signal with respect to the main XMCD peak. While not crucial for the subsequent analysis in this paper, these peaks represent hybridization of the $s$ states in Cu and Co. Close examination of the Cu DOS~\cite{ebertPRB} reveals that these satellite peaks are dominated by 2$p\rightarrow$4$s$ transitions in Cu, and it is also known that the 4$s$ states in Co are oppositely spin polarized compared to the 3$d$ states in Co, as observed in~\cite{obrienJAP} and calculated in~\cite{erikssonPRB,wuJAP}. This 4$s$ magnetization is much weaker than the 3$d$ contribution, but still induces an oppositely polarized 4$s$ magnetic contribution in Cu.

%\begin{figure}
%\includegraphics[width=0.50\textwidth]{fig2}% Here is how to import EPS art
%\caption{\label{fig:2} (color online) (a) A comparison of the XAS and XMCD spectra in Co/Cu alloys with 60\% Cu and 10\% Cu using two single spectra. The 10\% Cu shows both significantly narrowed XAS peak and a blue-shifted XMCD magnetic peak, which now closely overlaps with the XAS peak energy. Spectra energies are presented as the energy difference from the XAS L$_3$ peak. XAS and XMCD spectra were scaled to the same peak height to emphasize the other differences within these spectra. (b) Peak positions from the XAS/XMCD spectra for all four different alloy types. Each point is an average over all spectra taken on samples with that particular alloy concentration. No XMCD point is presented for 100\% Cu as there is no magnetization in pure Cu metal.}
%\end{figure}

\subsection{XAS Peak Position and Shape}

\begin{figure}
\centering
\includegraphics[width=0.49\textwidth]{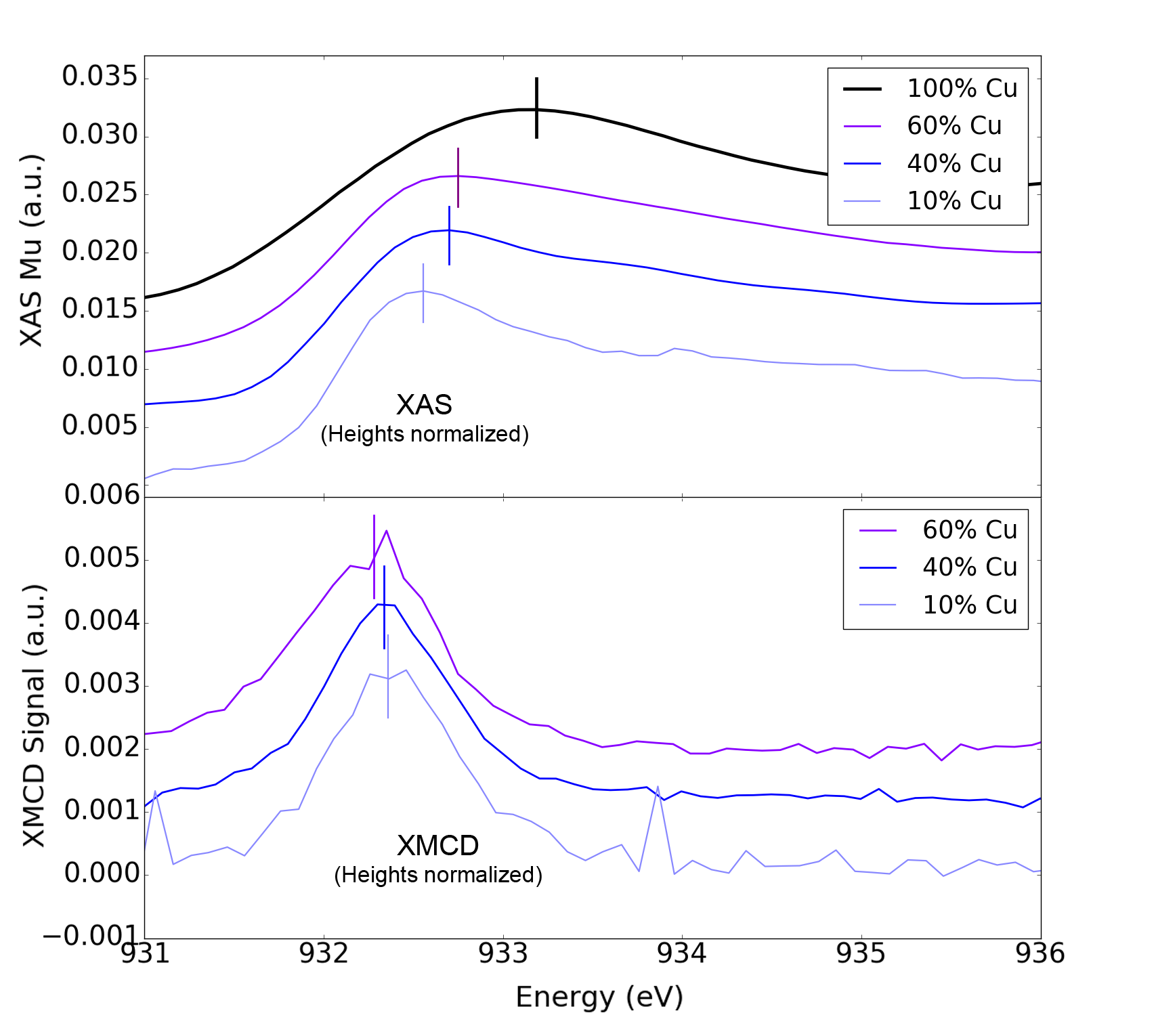}
\caption{\label{fig:2} (color online) %{\bf Please choose colors that are easier to distinguish, and you need to add the scaling factor to each XMCD spectrum in this figure. It took me a while to remember that these spectra were renormalized and thus the inset did not seem to make sense} 
Height-normalized spectra showing the dependence of the L$_3$ peak positions for both absorption (top) and magnetization (bottom), with spectra scaled to be the same amplitude. Solid vertical lines mark where the peak maximum was determined for these spectra. XAS spectra are clean enough for a simple maximum to be taken, but XMCD spectra are noisier and the peak maxima are determined as the center of the Gaussian-shaped XMCD spectrum. We scale all figures to have the same height, but for reference we also plot the unscaled XMCD signal in Figure \ref{fig:3}.} %{\bf I think we need to go a little bit deeper here. We can actually give an approximation for the Cu moment from the "delte mue t" observed e.g. at 40 per cent Cu, similar to what we do in Ni. Then we can use this to scale the data in the inset. If I remember it correctly I estimated the moment per Cu atom in the 40 per cent alloy to be 0.05 Bohr Magneton and I can check this one more time once I get back.
%I also think that your figure 3 should be the last one in the paper meaning figure 4, but with a second (bottom) panel showing the moments (that means the rescaled inset of the figure) and the FWHM inset from the next figure.}
\end{figure}

The top panel of figure~\ref{fig:2} shows magnetization averaged XAS spectra, where the spectra maxima have been marked with vertical solid lines. Spectra have been normalized to have approximately the same height for more easy comparison of peak positions. The observed spectra exhibit an opposing shift in peak position between the XAS and XMCD spectra. Peak positions at the L$_3$ edge were found using simple maxima of the normalized spectra; as a consistency check, we also subtract a linear interpolant that puts the pre-resonant and post-resonant portions of the spectra on the same level, and find that the relative shift is still consistent. For XMCD spectra, maxima from Gaussian fits of the spectra were used instead of the pure point-wise maxima to account for noise that would make the latter unreliable measures of the true feature maxima. Altogether we observe that the difference in photon energy between the maximum of the XAS peak and the XMCD peaks is continuously reduced the higher the Co concentration is or the more diluted the Cu atoms are within the Co matrix. A simple argument for this observation is that that the maximum DOS for Co d-orbitals is at the Fermi level, while for Cu it is about 0.5 eV above the Fermi level. Hence, in bulk Cu the Cu L-edge XAS peaks appears 0.5eV above E$_\textrm{F}$, while the XMCD appears at the maximum overlap between the Co  and Cu d-orbitals which is closer to E$_\textrm{F}$. With increased Co concentration and increased hybridization between Co and Cu the energy of the maximum in the Cu d-orbital DOS moves closer to E$_\textrm{F}$  \\

%The upper panel shows a comparison of Co rich (10\% Cu) and Co poor (40\% Cu) samples. Both spectra are normalized so that their peak heights are the same and are plotted with relation to the position of the XAS peak. We observe a marked narrowing of the L$_3$ XAS peak and also shifts in the satellite peaks post-edge, which will be discussed more thoroughly later in this paper. Most importantly,  In Figure~\ref{fig:2}b we show this shift on an absolute energy scale for both the XAS and XMCD peaks for all measured samples, demonstrating that the XMCD peak can shift up to 0.5eV.

%\begin{figure}
%\includegraphics[width=0.50\textwidth]{fig3}% Here is how to import EPS art
%\caption{\label{fig:3} (color online) (a) XMCD (magnetization) amplitude as a function of alloy Cu concentration. Values reported are for pure intensity differences for transmission with external applied field in opposite directions. (b) XAS peak FWHM as a function of alloy Cu concentration. Dotted lines mark the limit at a Cu concentration of 0\%.}
%\end{figure}

\begin{figure}
\centering
\includegraphics[width=0.5\textwidth]{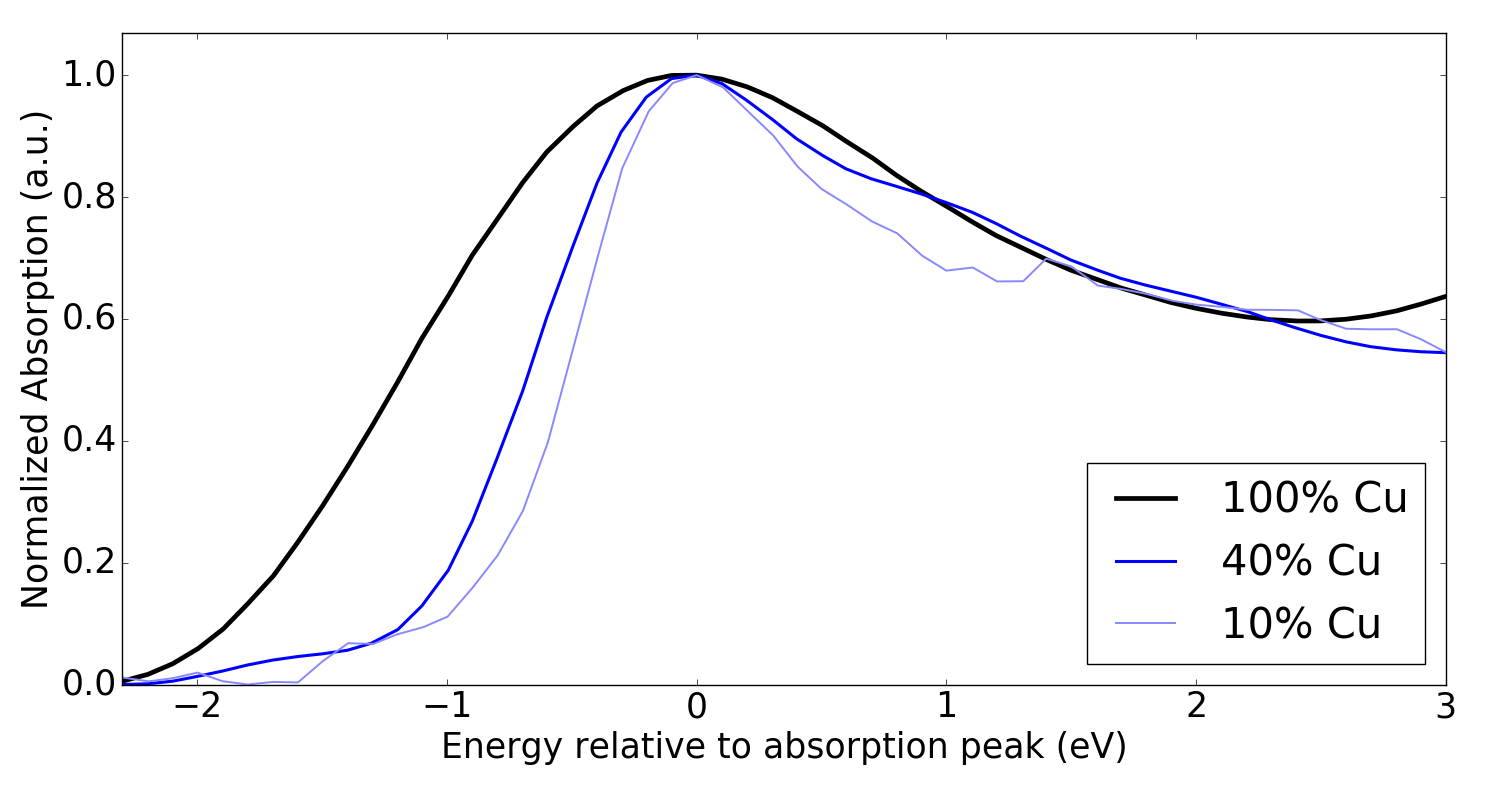}
\caption{\label{fig:3} (color online) Width comparisons of spectra at different alloy concentrations. Spectra are normalized to exhibit the same peak maximum value. The spectral width across the Cu L$_3$ edge can be seen to narrow in the raw spectra with increasing Co concentration.}
\end{figure}
In Figure~\ref{fig:3} we show how the relative width of the XAS spectra depends on the Cu concentration. The raw XAS spectra are plotted (with L$_3$ edge maxima aligned to each other) to emphasize that the narrowing effect is clearly visible. Quantitatively, the relative width is the ratio of the integral areas under the primary L$_3$ peaks of the XAS spectra for the alloy sample and the pure Cu sample. This method is mathematically equivalent to scaling the horizontal widths of the spectra by increasing $\alpha>1$ until they optimally overlap with the pure Cu spectra (in the least squares sense) and then taking the relative width to be $1/\alpha$. Quantitative consistency and robustness are maintained by repeating this calculation for regions of varying width symmetric around the peak, which also allows us to assign confidence intervals around these width measurements. Areas asymmetric with respect to the edge peak can also be used but do not result in appreciably different results. The quantitatively obtained FWHM of the XAS spectra are shown in the next section in Figure \ref{fig:4}. However, a qualitative observation can already be made here, namely that increased dilution of Cu in the Co matrix leads to decreased line width caused by increased localization of the Cu d-electrons. The FWHM of the XMCD peaks was also measured and was found to remain approximately constant. This is consistent with the fact that the magnetic signal in Cu is reflective of only emergent $d$-orbital character, which carries with it a well-defined lineshape.\\

\indent XAS spectra are also influenced by less localized continuum $s-p$ states, as reflected by Figure~\ref{fig:2} and Figure~\ref{fig:3}, and so we observe a changing lineshape due to a progression towards more localized $d$ states as a result of $d-d$ hybridization with the nearby Co atoms. This also indicates that the Cu atoms are evenly distributed throughout the matrix rather than clustered. In that case the Cu XAS spectrum would not show significant changes upon dilution. This narrowing is similar to those observed in~\cite{nilssonPRB} for a buried Cu multilayer, but here we see the entire trend for intermediate amounts of Co - the XAS width linearly narrows to a final value that is approximately 80\% the width of pure Cu L$_3$ peak. In conclusion, the interpretation of these observations is rather straightforward; in the limit of high Co concentration the coordination of Cu decreases and remaining Cu approaches its atomic-like behavior; Cu 3$d$-states are no longer spread across the entire film but tightly localized. This also indicates that the Cu atoms in our samples are evenly distributed throughout the alloyed sample, and are not forming Cu clusters.
\\ \linebreak

\subsection{Summary of Experimental Results}

\begin{figure}
\centering
\includegraphics[width=0.5\textwidth]{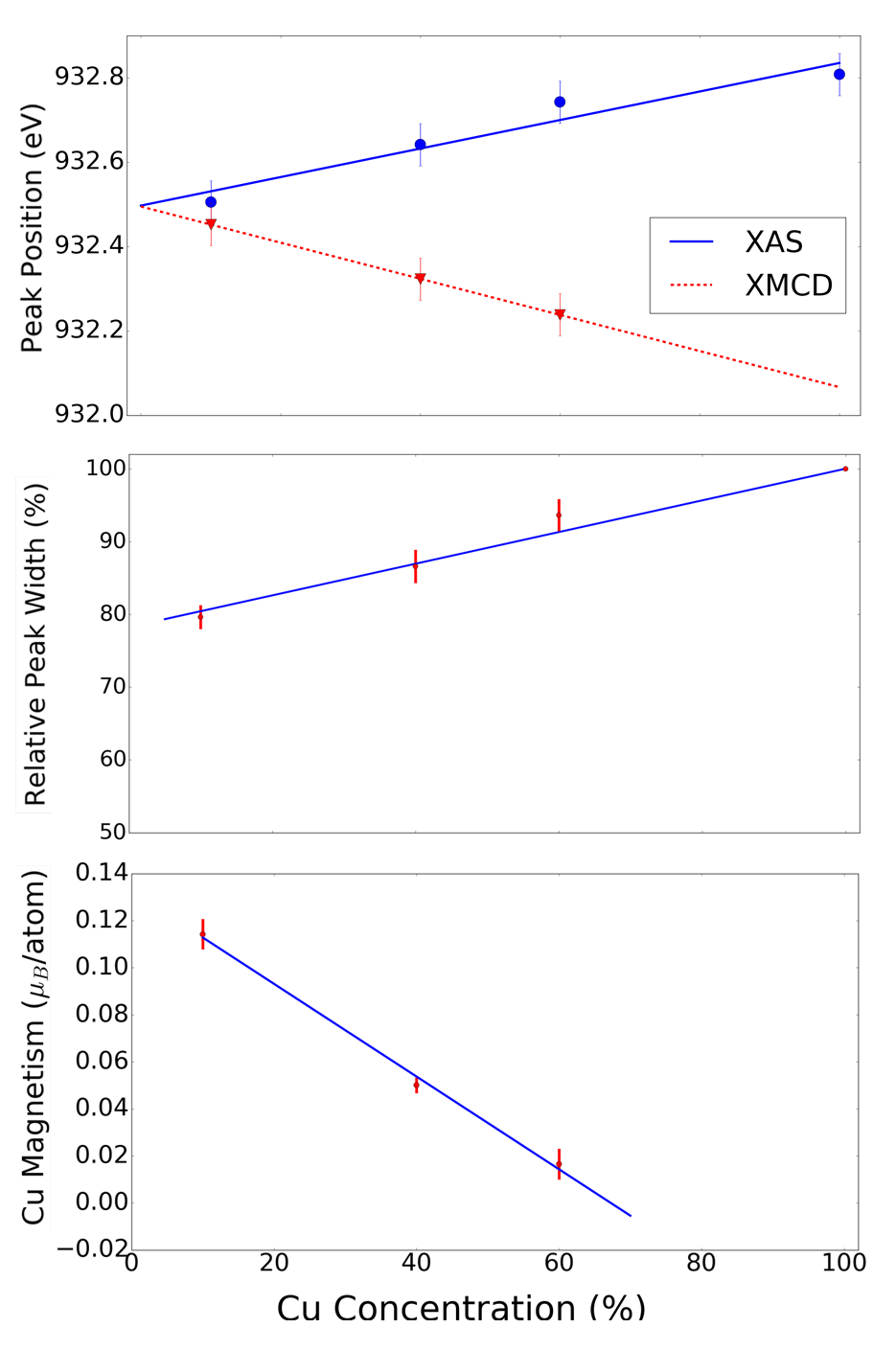}
\caption{\label{fig:4} (color online) Top: Peak positions from the XAS/XMCD spectra for all four different alloy types, averaged over all data. Each point is an average over all spectra taken on samples with that particular alloy concentration; the trend is thus more visually linear than that presented in Figure ~\ref{fig:2}, which showed single raw spectra. No XMCD point is presented for 100\% Cu as there is no induced static magnetization in pure Cu metal. Middle: The relative spectral widths to the pure copper spectra, defined as a ratio of areas under the L$_3$ absorption edges. Bottom: The magnetic moment per Cu as estimated from the peak XMCD value.
}
\end{figure}

We now summarize our experimental findings in figure~\ref{fig:4}. The top panel shows the position of the XAS and XMCD signal peak for alloyed samples at varying relative Cu concentrations, averaged over all spectra taken of that alloy. We find that with less Cu, both the XAS and XMCD peaks shift in opposite directions, with the XMCD peak shifting significantly to higher energies with respect to the XAS peak for a total relative shift of $\sim 0.7$eV (This assumes that the XMCD line can be extrapolated linearly to the pure Cu limit, even though there is no XMCD signal in that limit). In the limits of almost pure Cu, the XAS peak appears about 0.7eV above the Fermi energy (the maximum of the unoccupied Cu d-DOS) and XMCD would appear right at the Fermi energy from hybridization with diluted Co atoms.. The second panel shows that the FWHM of the Cu XAS L-edge spectrum continuously decreases with increased amount of Co, indicating that the Cu d-orbitals become more and more localized when diluted within the Co matrix, due to increased hybridization of Cu and Co d-orbitals. The last panel shows that for the case of highly diluted Cu a moment of the order of up to 0.1 Bohr Magneton  can be observed if Cu atoms are fully emerged in a Co matrix. To estimate the magnetic moment we use the approximation that based on the XMCD sum rules the XMCD intensity scales with the ratio of moment per hole in the 3d orbitals. Comparing the XMCD signal obtained for Cu with the XMCD signal commonly observed in Co, Ni or Fe allows us to estimate the size of the atomic moment in Cu. It should be noted that this estimate of the atomic moment will be of the correct order of magnitude, but it may over- or underestimate the moment by a single digit factor. Note that we observe a linear decrease of the induced moment with decrease of Co concentration. However, the Cu XMCD already vanishes around 70$\%$ Cu at room temperature, either due to formation of an antiferromagnetic or paramagnetic alloy at that concentration, due to the fact that the average separation of Co atoms is getting larger.

\subsection{Discussion}

\indent XAS core-level shifts of similar magnitude have been previously observed, for example in Cu/Ni multilayers~\cite{karisPRB} and Cu oxide systems~\cite{grioniPRB}, and are generally ascribed to changes in chemical environment. However, such effects have not been seen before in XMCD (previous studies did not magnetize their samples) as we have seen them here. We can explain this effect by noting that in the limit of high copper concentration, shifts in the Cu 3d states due to hybridization will have the most impact near $E_F$, as occupancy near $E_F$ are affected the most strongly by these shifts. We know that the Fermi energy $E_F$ in pure Cu lies at the inflection point of the XAS spectrum~\cite{roopaliPRL, grioniPRB}, whereas the Co XMCD features are centered precisely at the maximum of the L$_3$ resonance~\cite{chenPRL}. Thus, with increasing Co hybridization within the Cu orbitals, we observe an eventual alignment between the XAS and XMCD features within Cu. The XAS shift is towards lower energies, which is consistent with the chemical shifts as described in studies like ~\cite{karisPRB}, while the XMCD shift is a newly observed phenomenon and towards higher energies, consistent with the expectation that XMCD features should align with XAS features in typical 3d ferromagnets. Note that this also gives a natural interpretation to the limit in Figure~\ref{fig:2} for the pure Cu limit; the magnetic signal position approaches the Cu L$_3$ inflection point, even though it is not observable in this limit. One important consequence of the existence of this shift is that it opens up enticing possibilities in magnetic spectroscopy for Cu multilayers; in particular, it allows us to separate the interfacial magnetism with the bulk magnetism signal in Cu, as was demonstrated in~\cite{roopaliPRL}.

\indent The Figure~\ref{fig:2} inset shows the linear relationship between magnetic signal (XMCD amplitude) and Co alloy concentration. The movement of peak positions in Figure~\ref{fig:2} seems linear, although one may expect that the behavior is not linear but rather asymptotic with increasing Cu concentration. However, due to the vanishing magnitude of the XMCD effect for large Cu (small Co) concentrations, a reliable quantitative measurement is not possible in that regime. What seems indisputable is that the moment per Cu atom increases with increasing Co concentration, which indicates again that the Cu is dispersing close to uniformly within the Co matrix (i.e. Cu atoms are not forming clusters).\\ \linebreak

\section{\label{sec:level1}Conclusions}

We present a detailed X-ray spectroscopic studies of Cu in Co/Cu alloys, and develop a consistent picture of emergent localized magnetization and increased 3$d$ character in Cu. Not only do we see an enhanced magnetic signal in Cu with increased Co, but we also see significant changes in peak positions, widths, and general shapes of the features in our spectra, all of which are consistent with an increased 3$d$ empty state DOS in Cu due to $d-d$ hybridizing with neighboring Co and increasing $d$-state localization. Ultimately, our measurements show that electronic behavior at the interface between a ferromagnet and nonmagnet is dramatically different than behavior in the metallic bulk, and provide a detailed look at the rich interface physics that will take center stage as efforts in magnetism push even further into the nanoscale.

\section{\label{sec:acknowledgements}Acknowledgements}

Work at University of California San Diego including materials synthesis and characterization and participation in synchrotron measurements was supported by U.S. Department of Energy (DOE), Office of Basic Energy Sciences (BES) under Award No. DESC0003678. The research at NYU was supported by Grant No. NSF-DMR-1610416.
%\nocite{*}

\bibliography{CoCu_hendrik}% Produces the bibliography via BibTeX.

\end{document}